\documentclass[12pt]{article}
\usepackage{graphicx}
\usepackage{hyperref}

\newcommand\pubnumber{DPF2015-66}
\newcommand\pubdate{\today}

\def\nebraska{Department of Physics and Astronomy\\
University of Nebraska-Lincoln, Lincoln, NE 68588-0299}

\def\Title#1{\begin{center} {\Large #1 } \end{center}}
\def\Author#1{\begin{center}{ \sc #1} \end{center}}
\def\Address#1{\begin{center}{ \it #1} \end{center}}

\newcommand\pubblock{\rightline{\begin{tabular}{l} \pubnumber\\
         \pubdate  \end{tabular}}}
\newenvironment{Abstract}{\begin{quotation}  }{\end{quotation}}
\newenvironment{Presented}{\begin{quotation} \begin{center} 
             PRESENTED AT\end{center}\bigskip 
      \begin{center}\begin{large}}{\end{large}\end{center} \end{quotation}}
\def\Acknowledgments{\bigskip  \bigskip \begin{center} \begin{large}
             \bf ACKNOWLEDGMENTS \end{large}\end{center}}
\def\beq{\begin{equation}}
\def\eeq#1{\label{#1}\end{equation}}
\def\eeqn{\end{equation}}

\def\beqa{\begin{eqnarray}}
\def\eeqa#1{\label{#1}\end{eqnarray}}
\def\eeqan{\end{eqnarray}}

\let\bar=\overbar

\def\Dslash{\not{\hbox{\kern-4pt $D$}}}
\def\dslash{\not{\hbox{\kern-2pt $\del$}}}

\def\msb{{\bar{\ssstyle M \kern -1pt S}}}

\begin{document}
\begin{titlepage}
\pubblock

\vfill
\Title{Search for associated production of a Higgs boson with a single top quark}
\vfill
\Author{Kenneth Bloom}
\Address{\nebraska}
\vfill
\begin{Abstract}
  We present a search for the production of a Higgs boson in association
  with a single top quark ($tHq$), using data samples collected by the CMS
  detector in $pp$ collisions at center-of-mass energy of 8~TeV corresponding
  to integrated luminosity of 19.7~fb$^{-1}$. The search exploits a variety of
  top quark and Higgs boson decay modes resulting in final states with
  photons, bottom quarks, or multileptons, and employs a variety of machine
  learning techniques to maximize the sensitivity to the signal. The
  present analysis is optimized for a scenario where the Yukawa coupling
  has sign opposite to the standard model prediction, which would result in
  a large enhancement of the signal cross section. Results for individual
  final states and the combined results will be presented.
\end{Abstract}
\vfill
\begin{Presented}
DPF 2015\\
The Meeting of the American Physical Society\\
Division of Particles and Fields\\
Ann Arbor, Michigan, August 4--8, 2015\\
\end{Presented}
\vfill
\end{titlepage}
\def\thefootnote{\fnsymbol{footnote}}
\setcounter{footnote}{0}

\section{Introduction}

Since the discovery of a Higgs boson, we have all become quite familiar
with the four favorite Higgs production modes: direct production through
gluon fusion, vector boson fusion, associated production with a $W$ or $Z$
boson, and associated production with a $t\bar{t}$ pair.  The measurements
of the rates for these production processes generally agree well with the
predictions of the standard model~\cite{bib:higgscouplings}.  But there is
another production mode to consider: the associated production of a Higgs
boson with a single top quark through a $t$-channel electroweak
interaction.  (The final state is denoted as $tHq$ because of the spectator
quark that appears.)  The cross section for this process is expected to be
about a thousand times smaller than that for gluon fusion, with
$\sigma(tHq)$ = 18.3~fb.  Why even bother looking for something with such a
tiny rate in the Run~1 LHC data?

\begin{figure}[htb]
\centering
\includegraphics[height=1.5in]{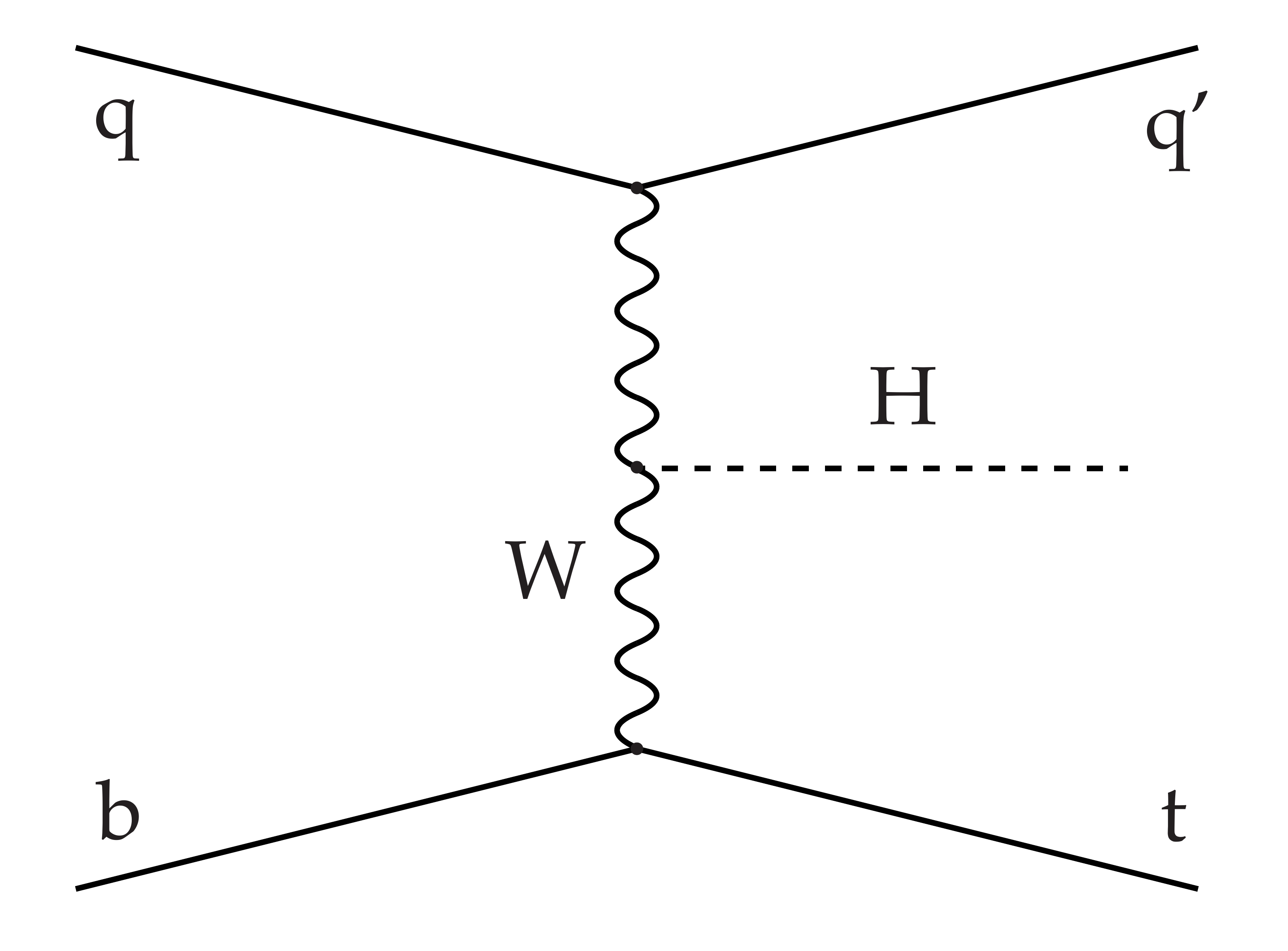}
\includegraphics[height=1.5in]{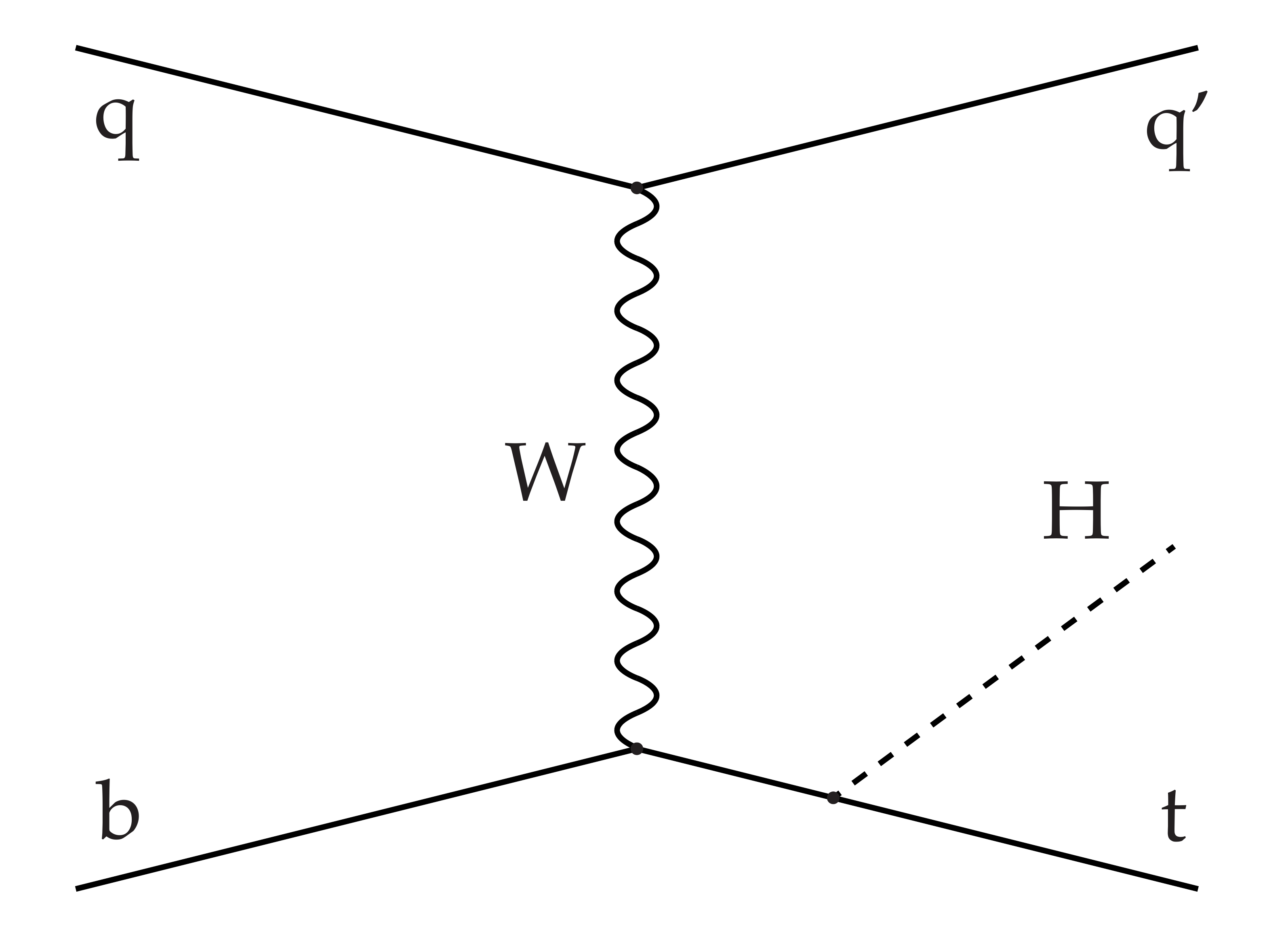}
\caption{Dominant Feynman diagrams for the production of $tHq$ events: the
  Higgs boson is typically radiated from the heavier particles of the diagram,
  {\it i.e.} the $W$ boson (left) or the top quark (right).}
\label{fig:diagrams}
\end{figure}

The answer can be seen in the two diagrams for $tHq$ production shown in
Figure~\ref{fig:diagrams}.  The Higgs boson emerges from either an internal
$W$ boson
or an external top quark.  The cross section is small because of the destructive
interference between these two diagrams.  There is nothing deep about this
-- there is no symmetry that keeps the rate small -- but it is how the
calculation turns out.  However, should the sign of the top Yukawa coupling
be inverted ($C_t = -1)$, the destructive interference becomes
constructive, and the predicted cross section would increase by a factor of
about 13~\cite{bib:theory}.  While $C_t = -1$ is disfavored by the current
data, it is not completely eliminated, and other new physics such as
flavor-changing neutral current (FCNC)
processes could also enhance the rate for $tHq$ production.  Thus this
channel provides an interesting path to discovery through interference
effects.

The CMS Collaboration has completed four direct searches for the $tHq$
process, corresponding to different Higgs boson decay modes.  All four
search explicitly for the $C_t = -1$ process.  All take advantage of
top-quark leptonic decay, which provides a lepton, $b$ jet and missing
energy in the final state.  All have $t\bar{t}$ production as the most
significant background, as $t\bar{t}$ events (including $t\bar{t}H$)
contain many of the same elements as the the $tHq$ signal.  The four Higgs
final states explored are $\gamma \gamma$ (small branching ratio but very
pure, plus an enhancement in the $H \to \gamma \gamma$ branching fraction
due to $C_t = -1$); $WW$/$\tau\tau$ decaying to multiple leptons (large
branching fraction, but has non-prompt lepton backgrounds);
$\tau_{\mathrm{had}}\tau_\ell$ (similar issues to multileptons but even
smaller rate); and $b\bar{b}$ (largest branching ratio but very large
$t\bar{t}$ background).  The results of the four searches are combined
into a single limit on anomalous $tHq$ production.  The
$\gamma\gamma$~\cite{bib:ggpas}, $WW$/$\tau\tau$~\cite{bib:multileppas} and
$b\bar{b}$~\cite{bib:bbpas} results have previously been presented in
preliminary form; the $\tau_{\mathrm{had}}\tau_\ell$ result and the
combination were presented for the very first time at this conference.  The
remainder of this paper gives brief descriptions of each of the four
analyses and of the combination.

\section{The searches}

\subsection{$H \to \gamma \gamma$}

The $H \to \gamma\gamma$ analysis searches for a narrow di-photon resonance
with a mass near that of the Higgs boson,
$122 < m_{\gamma\gamma} < 128$~GeV.  In addition,
candidate events are required to have a single isolated electron or muon
candidate, one identified $b$ jet, and one forward jet; these are the decay
products of the single top and the spectator quark.

Background processes that produce real $H \to \gamma\gamma$ decays will
give events in the signal region of $m_{\gamma\gamma}$.  These are reduced
with a cut on a likelihood discriminant designed to distinguish $t\bar{t}H$
from $tHq$ production that is formed from various kinematic quantities in
the event.  Simulations are then used to estimate the rate.  The
contributing processes are $t\bar{t}H$ (0.03 + 0.05 events), $VH$ (0.01 +
0.01 events); all other Higgs production mechanisms give neglible
contributions.  The second number in the pairs listed reflects the
enhancement in $B(H \to \gamma\gamma)$ due to the assumed value of
$C_t = -1$.  All remaining backgrounds have a smooth shape in
$m_{\gamma\gamma}$ and can be estimated from the data using the sidebands
in that quantity.  The $m_{\gamma\gamma}$ spectrum is fit with an
exponential function, and higher-statistics control regions are used to
estimate systematic uncertainties on the fit.

\begin{figure}[htb]
\centering
\includegraphics[height=2.5in]{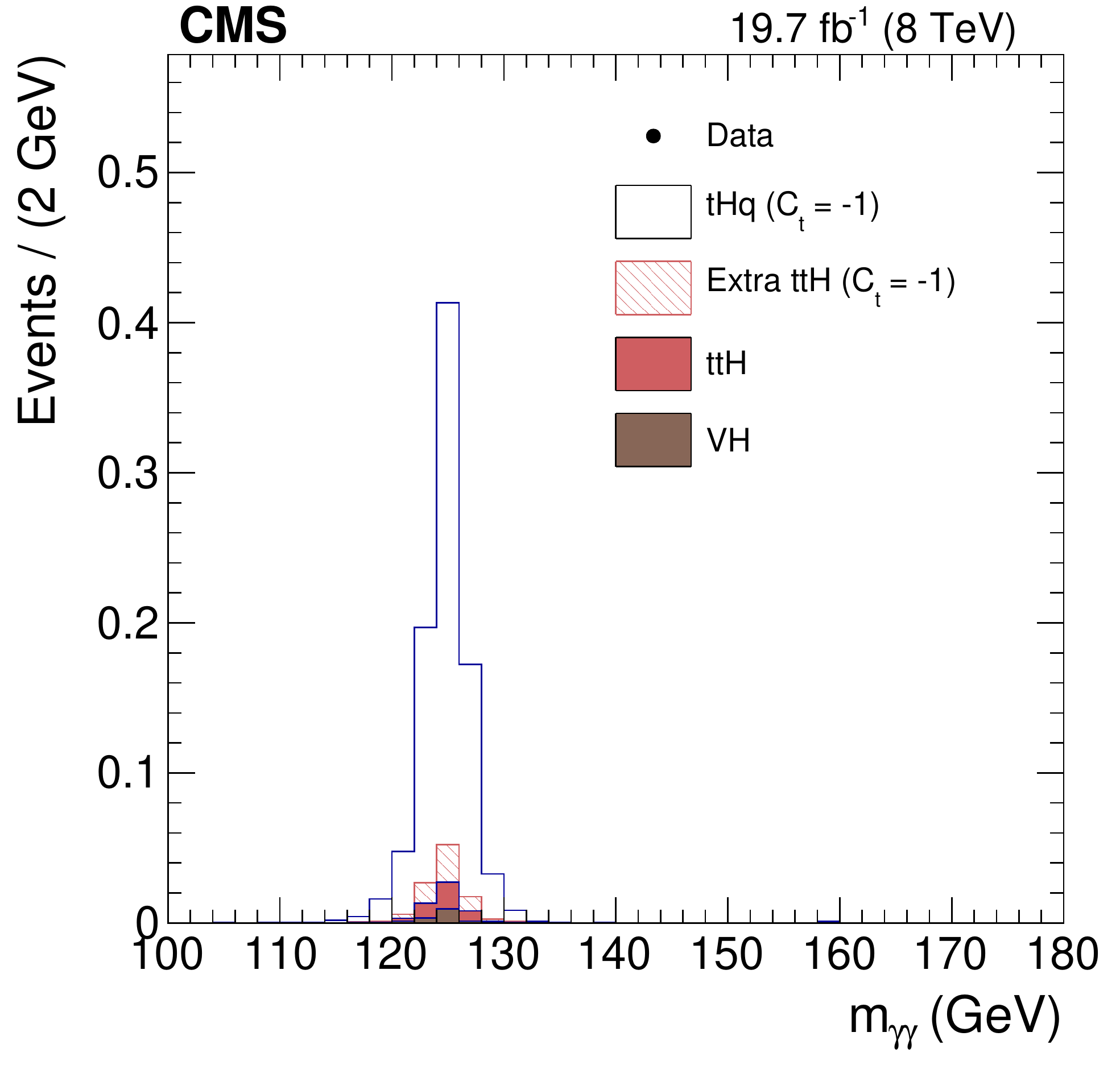}
\caption{Invariant mass of the diphoton system for events passing all event
  selection requirements. No events are observed after the cut on the
  likelihood discriminant.}
\label{fig:gg}
\end{figure}

The resulting $m_{\gamma\gamma}$ spectrum is shown in Figure~\ref{fig:gg}.
Zero events are observed in the sidebands, and zero events are observed in
the signal region also.  This results in a 95\% confidence level upper limit of $4.1 \times
\sigma_{tHq}(C_t = -1)$, which coincides with the expected limit.

\subsection{$H \to WW$/$\tau\tau$}

The multilepton final state targets cases where the Higgs boson decays to $WW$
which then decay to leptons (either $e$ or $\mu$), but also captures some
amount of $H \to \tau\tau$ with each $\tau$ decaying to $e$ or $\mu$.
Along with the semileptonic top decay, this gives three leptons in the
final state.  This allows for a trilepton search in the final states
$\mu\mu\mu$, $\mu\mu e$, $\mu ee$ and $eee$.  In addition to those leptons,
exactly one $b$-tagged jet, at least one forward non-$b$-tagged jet and
missing energy are required.  Events consistent with a $Z \to \ell\ell$
boson decay are vetoed.

A search involving same-sign dileptons will detect events in which one of
the $W$ bosons decays hadronically and the other decays leptonically.  A
same-sign pair ($e\mu$ or $\mu\mu$) is chosen to reduce backgrounds from
Drell-Yan processes.  In addition, events are required to have at least one
$b$-tagged jet, at least one central jet, and at least one forward
non-$b$-tagged jet.  Hadronic tau decays are explicitly rejected.

In all of the channels, about half of the background events come from
non-prompt leptons, most of which come from $t\bar{t}$ decays.  The rate
for this background is estimated with a ``tight-loose'' method, in which a
fake rate is measured in control samples, and then applied to events that have
similar kinematics to those in the signal sample but reside in
identification and isolation sideband regions.  There are also background
events due to misidentified lepton charge; the rate for this is estimated
from $Z$ boson events.  The signal events can be distinguished from other events
through a discriminating likelihood function that is formed from
information on forward activity, jet and $b$-jet multiplicity, and lepton
kinematics and charge.  The expected number of signal and background events
are 3.3 and 106 in the $e\mu$ sample, 2.2 and 53 in the $\mu\mu$ sample and
1.5 and 42 in the $3\ell$ sample.

%\begin{table}[t]
%\begin{center}
%\begin{tabular}{c|ccc}  \hline
%Prediction &  $e\mu$ &  $\mu\mu$ &  $3\ell$ \\ \hline
%Signal & 3.3 & 2.2 & 1.5\\
%Background & 106 & 53 & 42\\\hline
%\end{tabular}
%\caption{Predicted number of signal and background events in the
 % multilepton samples.}
%\label{tab:multilep}
%\end{center}
%\end{table}

\begin{figure}[htb]
\centering
\includegraphics[height=2in]{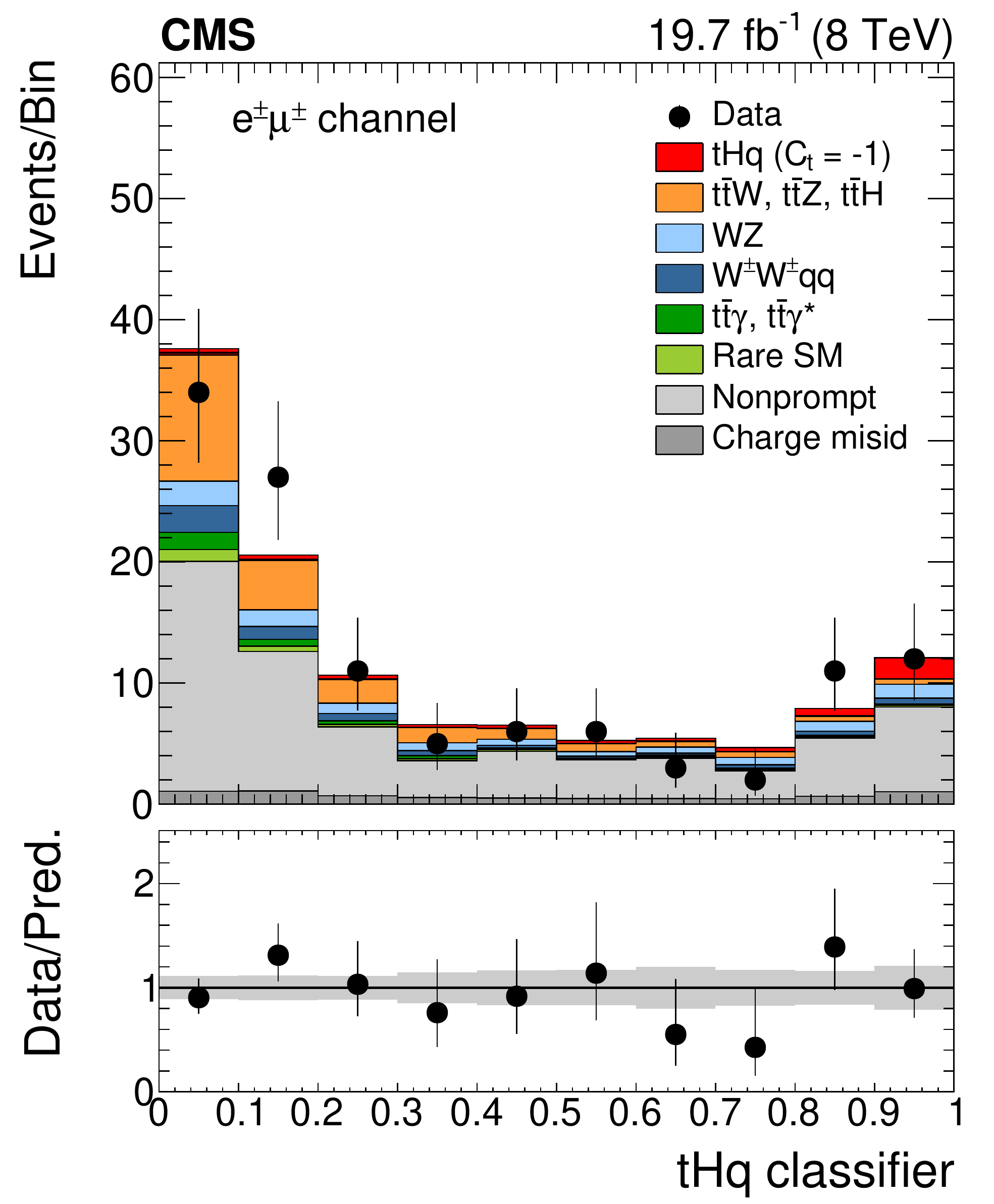}
\includegraphics[height=2in]{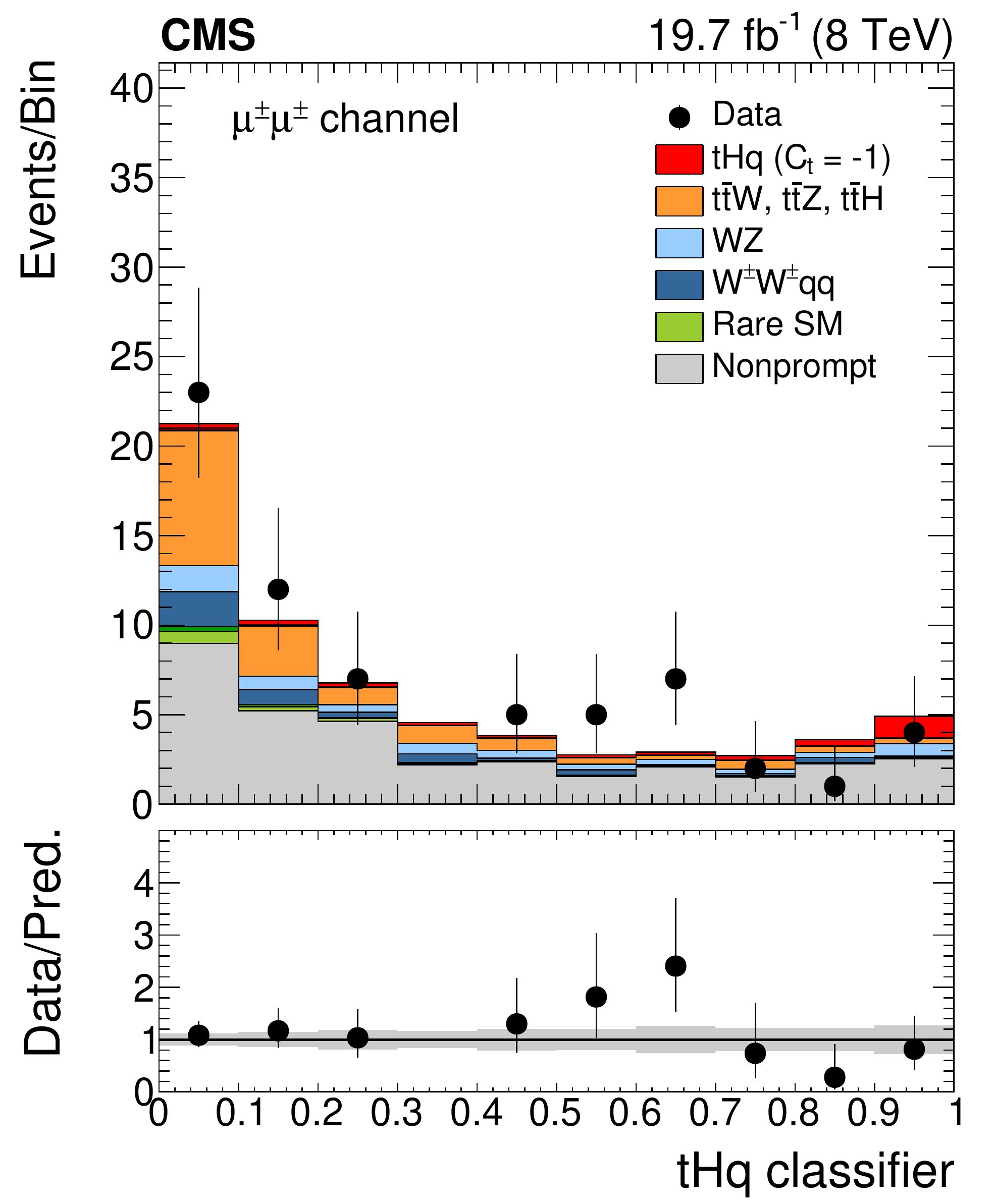}
\includegraphics[height=2in]{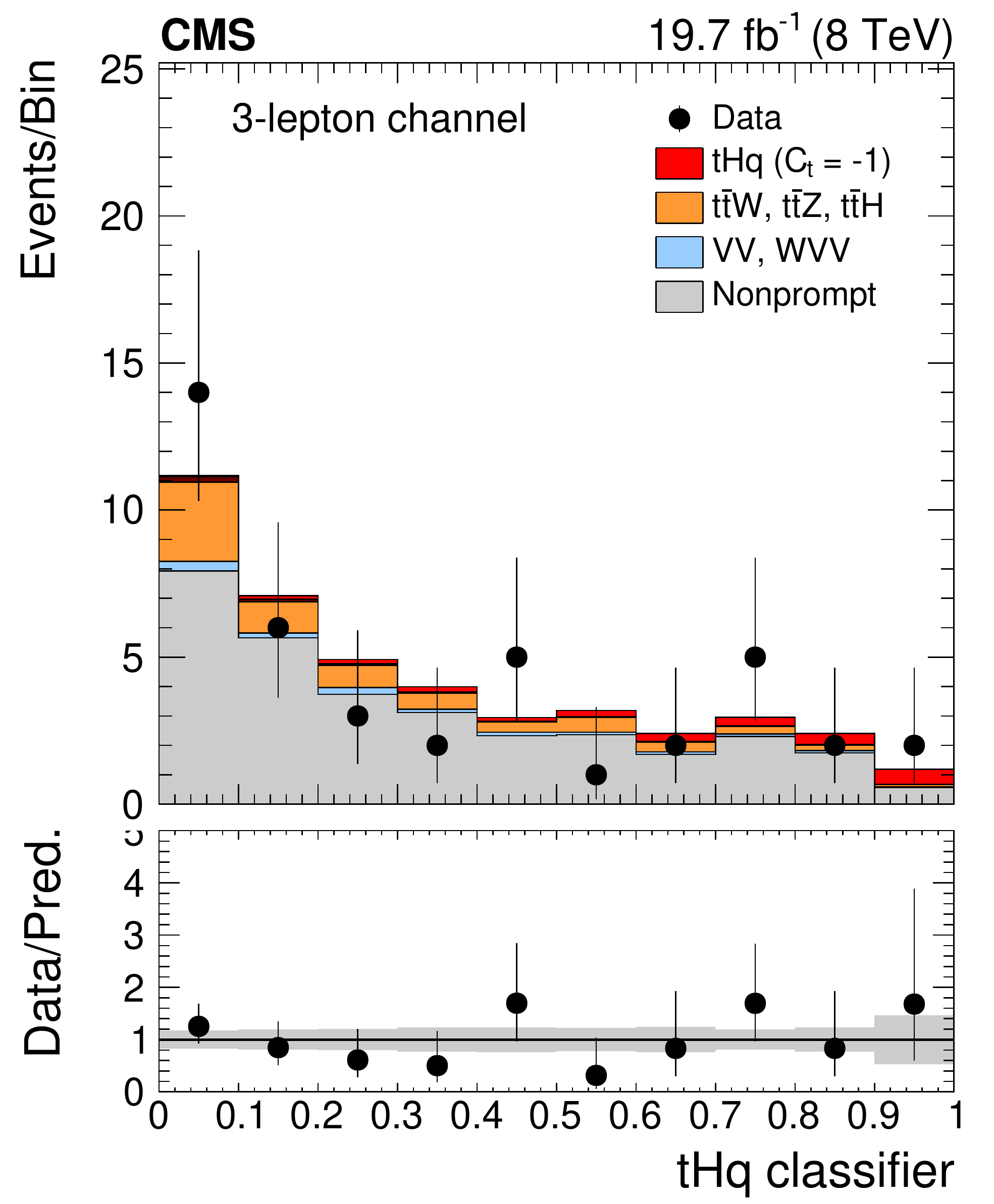}
\caption{Discriminating function output, for the $e\mu$ (left), $\mu\mu$
  (center), and trilepton channel (right).  In the box below each
  distribution, the ratio of the observed and predicted event yields is
  shown.  The gray band represents the post-fit systematic and statistical
  uncertainties.}
\label{fig:multilep}
\end{figure}

Figure~\ref{fig:multilep} shows the output of the discriminating function for
those data samples, along with the physics components of the sample
normalized to the results of a likelihood fit that is used to set a limit
on the production rate.  A combination of the results from the three
samples gives an expected upper limit of $5.0^{+2.1}_{-1.4} \times
\sigma_{tHq}(C_t = -1)$ at the 95\% confidence level, and an observed limit of
$6.7 \times\sigma_{tHq}(C_t = -1)$.  The largest systematic uncertainties
arise from the non-prompt lepton rate estimate.

\subsection{$H \to \tau_{\mathrm{had}}\tau_\ell$}

The search space for $tHq$ can be expanded further by including $H \to
\tau_{\mathrm{had}}\tau_\ell$ decays.  This search is similar to the
trilepton analysis, but incorporates a hadronically-decaying $\tau$ as one
of the three leptons.  The desired signature is a same-sign $e\mu$ or
$\mu\mu$ pair, with one lepton coming from top decay and the other from a
leptonic $\tau$ decay.  The analysis uses a multivariate technique to
evaluate the lepton isolation.  An isolated $\tau_{\mathrm{had}}$ and a
least one $b$-tagged jet are also required.  The largest background in the
analysis is from $t\bar{t}$ events with non-prompt leptons.  As with the
trilepton analysis, the contribution from non-prompt leptons is estimated
by determining a fake rate and applying it to a sideband sample.  And
similarly to the trilepton analysis, a Fisher discriminant is built to separate
the background processes, using the properties of the most-forward jet,
$b$-jet properties and other kinematic variables.  A sample with inverted 
$\tau_{\mathrm{had}}$ isolation is used for training and validation of the
discriminant.

\begin{figure}[htb]
\centering
\includegraphics[height=2.5in]{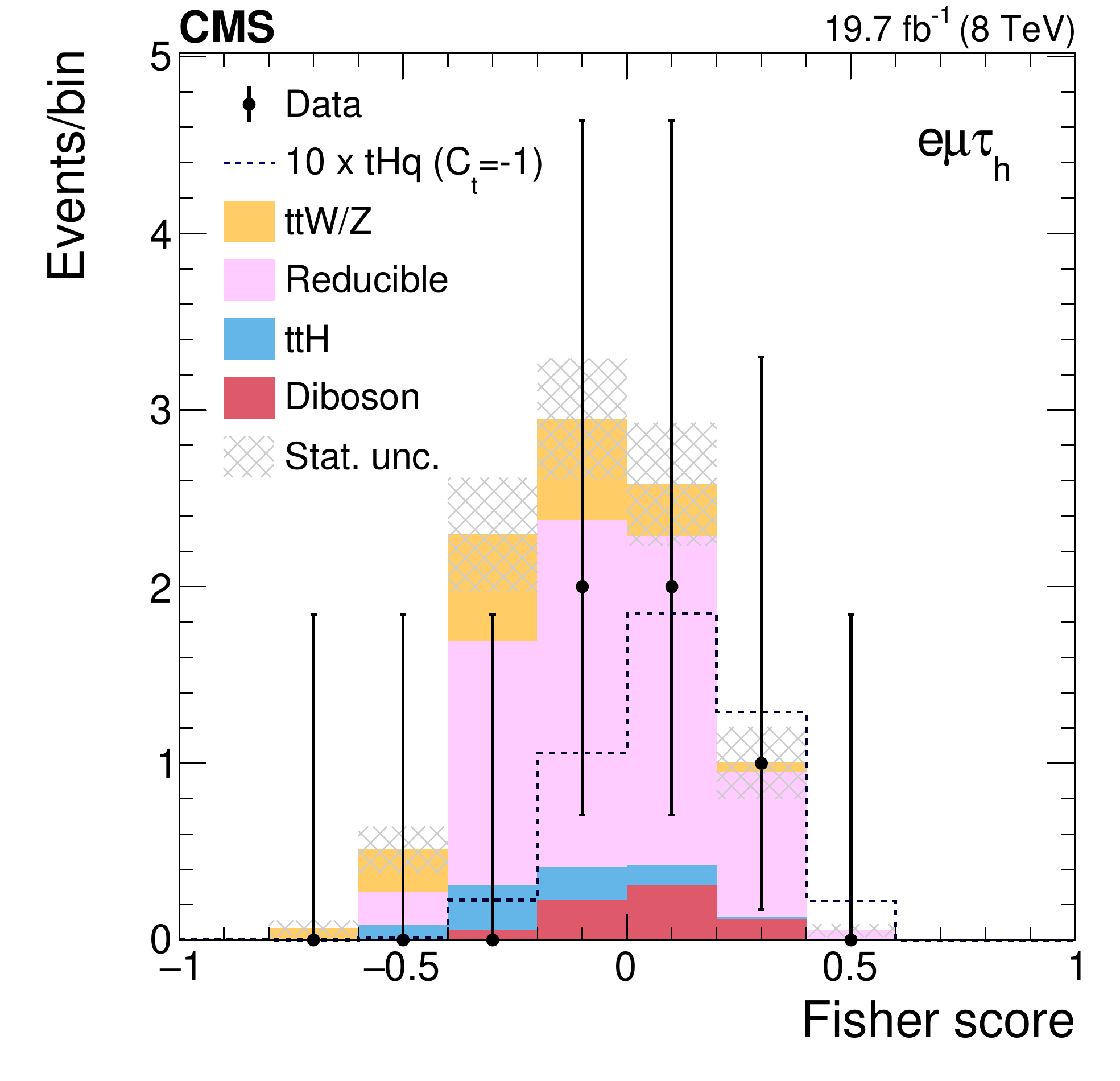}
\includegraphics[height=2.5in]{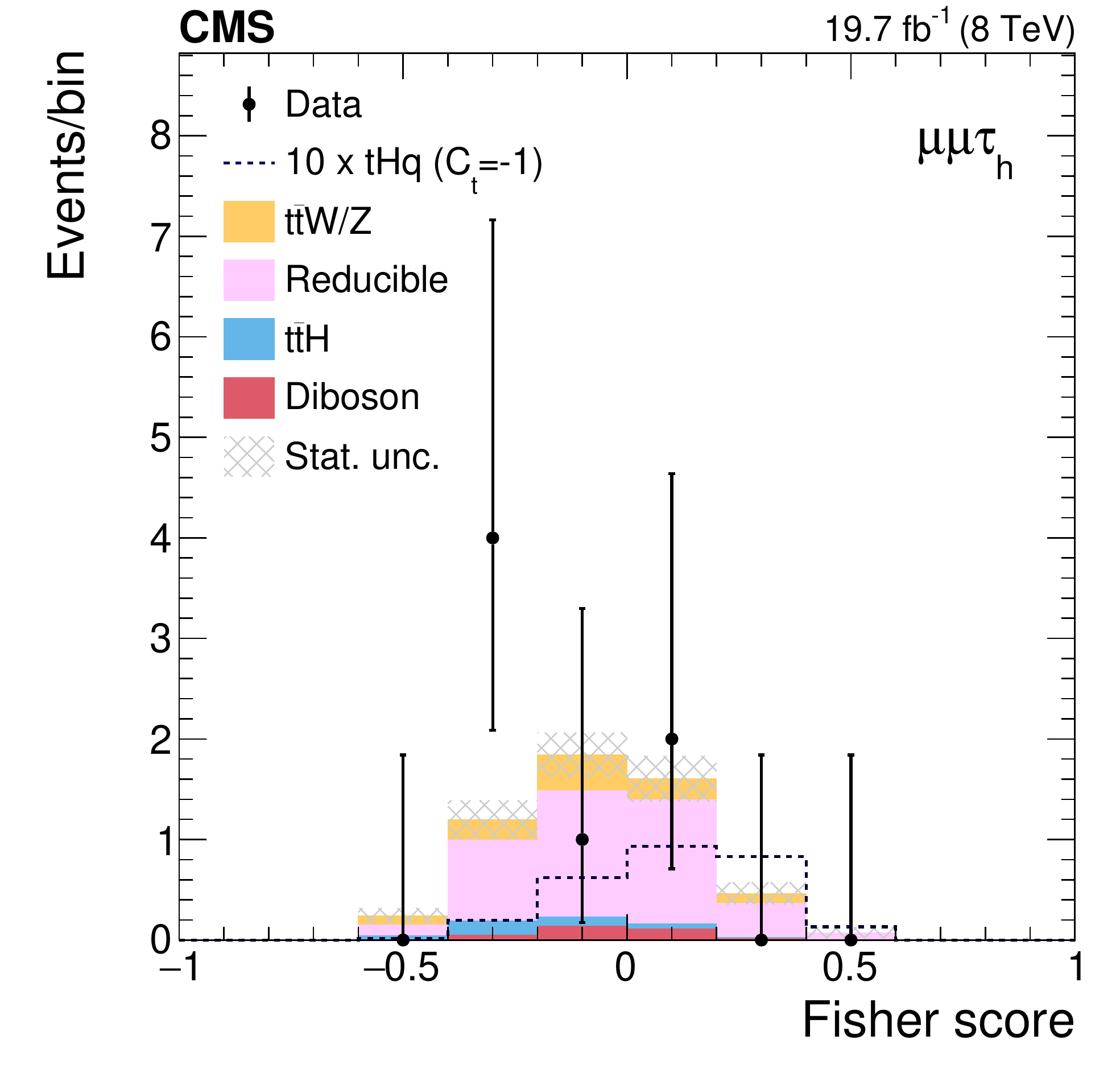}
\caption{Expected (histograms) and observed (points) distributions of the
  Fisher discriminant in the $e\mu\tau_{\mathrm{had}}$ channel (left) and
  $\mu\mu\tau_{\mathrm{had}}$ channel (right). The dashed line gives the
  expected contribution from the tHq signal ($C_t = −1$) case, multiplied
  by ten.}
\label{fig:tau}
\end{figure}

A total of 0.48 (0.30) signal events are expected in the
$e\mu \tau_{\mathrm{had}}$ ($\mu\mu\tau_{\mathrm{had}}$) sample, along with
9.5 (5.4) background events.  A total of 5 (7) events are observed in the
data.  The distributions of the Fisher discriminant values is shown in
Figure~\ref{fig:tau}.  A maximum likelihood fit of these distributions is
used to obtain limits on the production rate.  For these two samples, the
fit gives an expected upper limit of
$11^{+6}_{-4} \times \sigma_{tHq}(C_t = -1)$ at the 95\% confidence level, and
an observed limit of $9 \times\sigma_{tHq}(C_t = -1)$.  The largest
systematic uncertainties in the analysis come from the estimate of the
non-prompt lepton background rates, but the statistical uncertainties
dominate the limit that is obtained.

\subsection{$H \to b\bar{b}$}

The $H \to b\bar{b}$ decay mode provides a promising avenue for the $tHq$
search because of the large branching ratio for that decay.  The event
selection requires one isolated high transverse momentum ($p_T$) lepton
($e$ or $\mu$), missing energy from the neutrino in the $W$ boson decay, a
non-$b$-tagged jet that is either forward or higher in $p_T$, and three or
four $b$-tagged jets -- one from the top-quark decay, two in the Higgs
boson decay, and
possibly one more from gluon splitting in the initial-state partons.
Unfortunately, $t\bar{t}$ with extra heavy-flavor jets or mistagged light
jets can also produce such a final state.  The expected signal to
background ratio is 13/1900 in the 3$b$ sample and 1.4/66 in the 4$b$
sample.

The large number of jets in the final state leads to complications.  Each
of the jets needs to be assigned to parent quarks in final state.  A
multivariate discriminator based on event quantities such as invariant
masses, jet angular separations, jet $\eta$ and $p_T$ values and jet
charges is used to do the assignment.  The single best assignment of jets
to quarks is used as the reconstruction hypothesis.  This is done twice,
under two different assumptions of the final state -- $tHq$ signal and
$t\bar{t}$ background.  Once these reconstructions are performed, kinematic
quantities specific to each reconstruction are formed, and used to develop
another discriminator that distinguishes the two processes.  Templates in
that variable are then used to extract the $tHq$ signal fraction.  The
$t\bar{t}$ template is taken from simulation, allowing the amount of
additional heavy flavor quarks in those events to vary.  The simulation is
verified with a data-driven method that makes use of 2$b$ events to model
the $t\bar{t}$ kinematics in the signal sample and gives consistent
results.

\begin{figure}[htb]
\centering
    \includegraphics[scale=0.30]{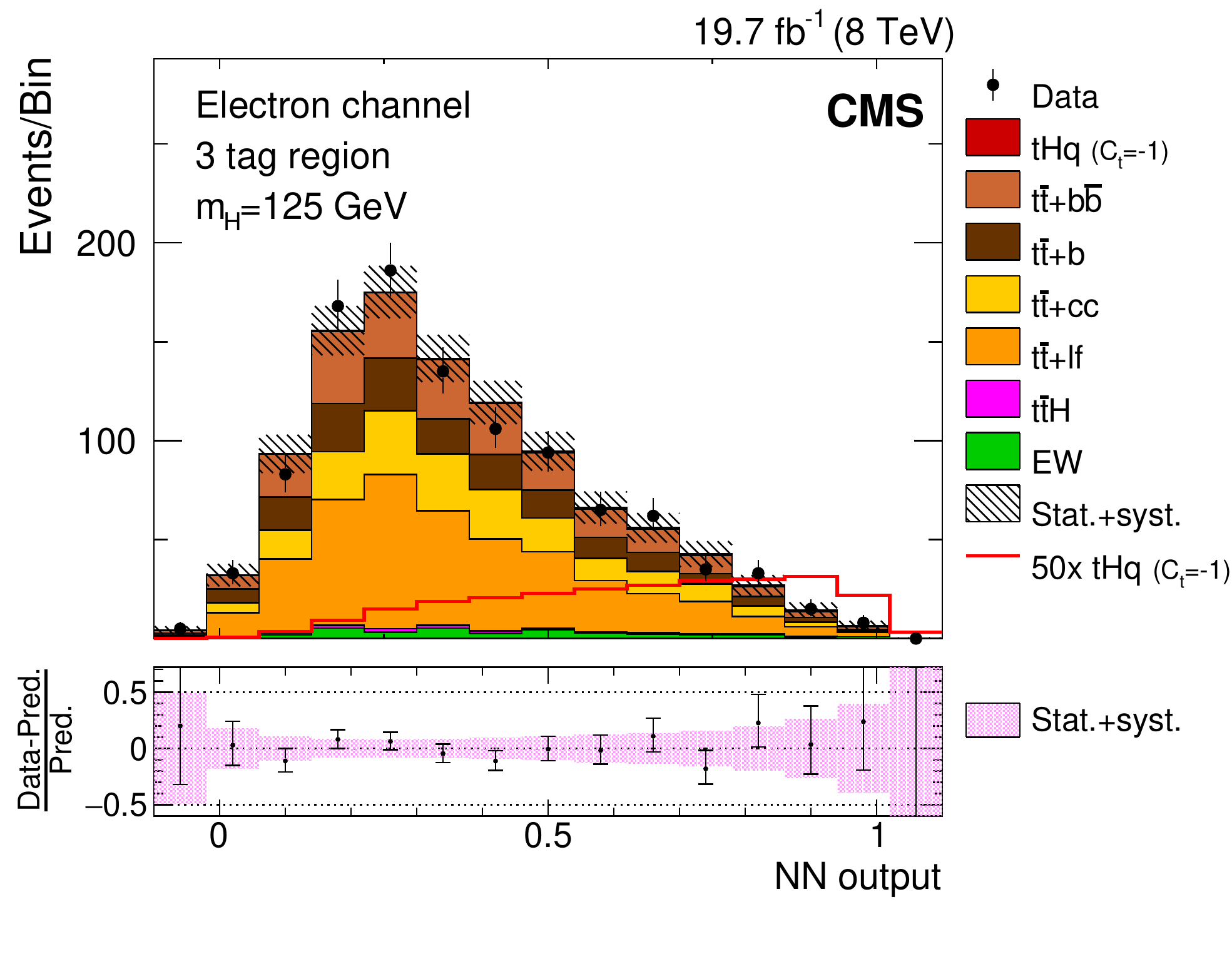}
    \includegraphics[scale=0.30]{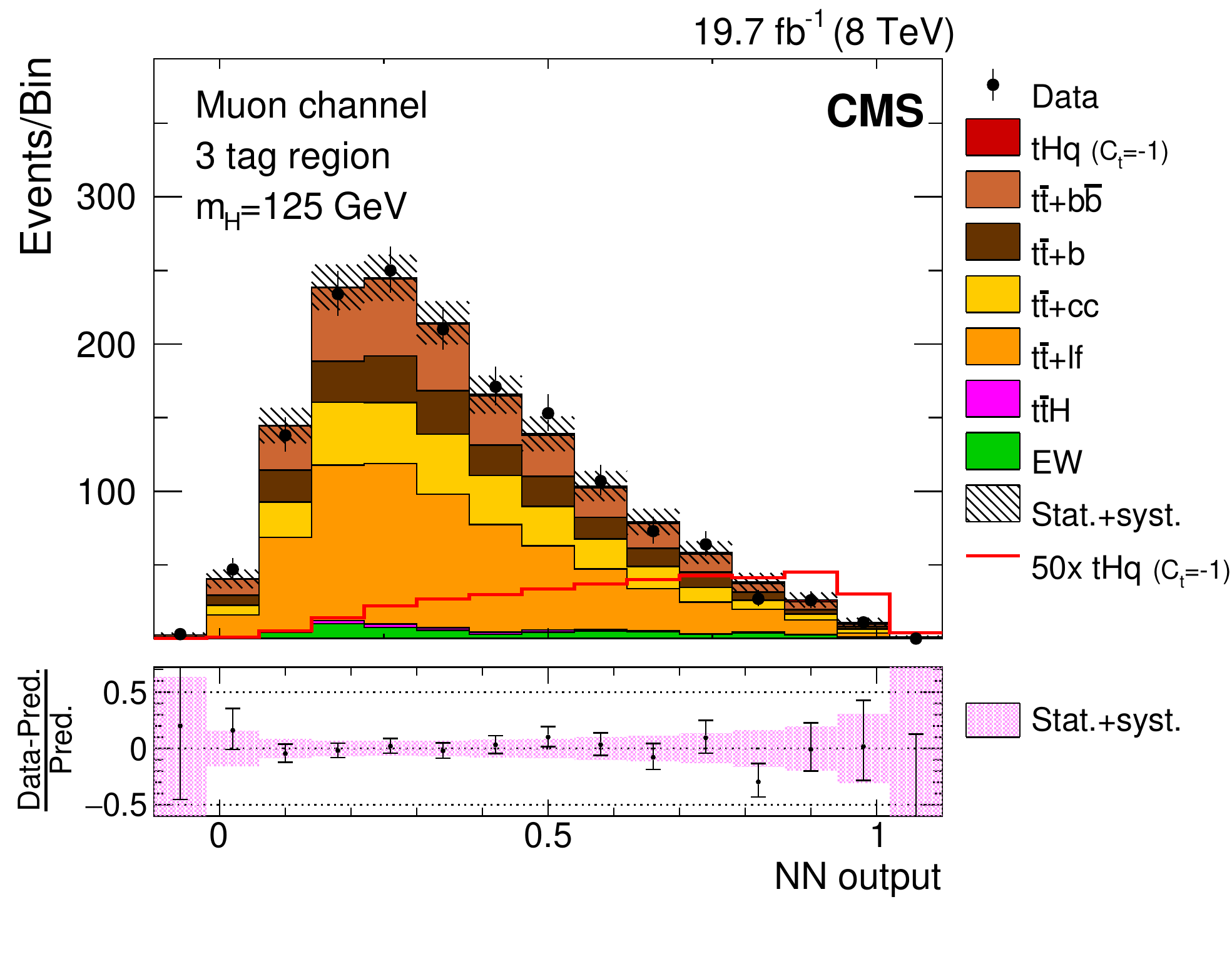}\\
    \includegraphics[scale=0.30]{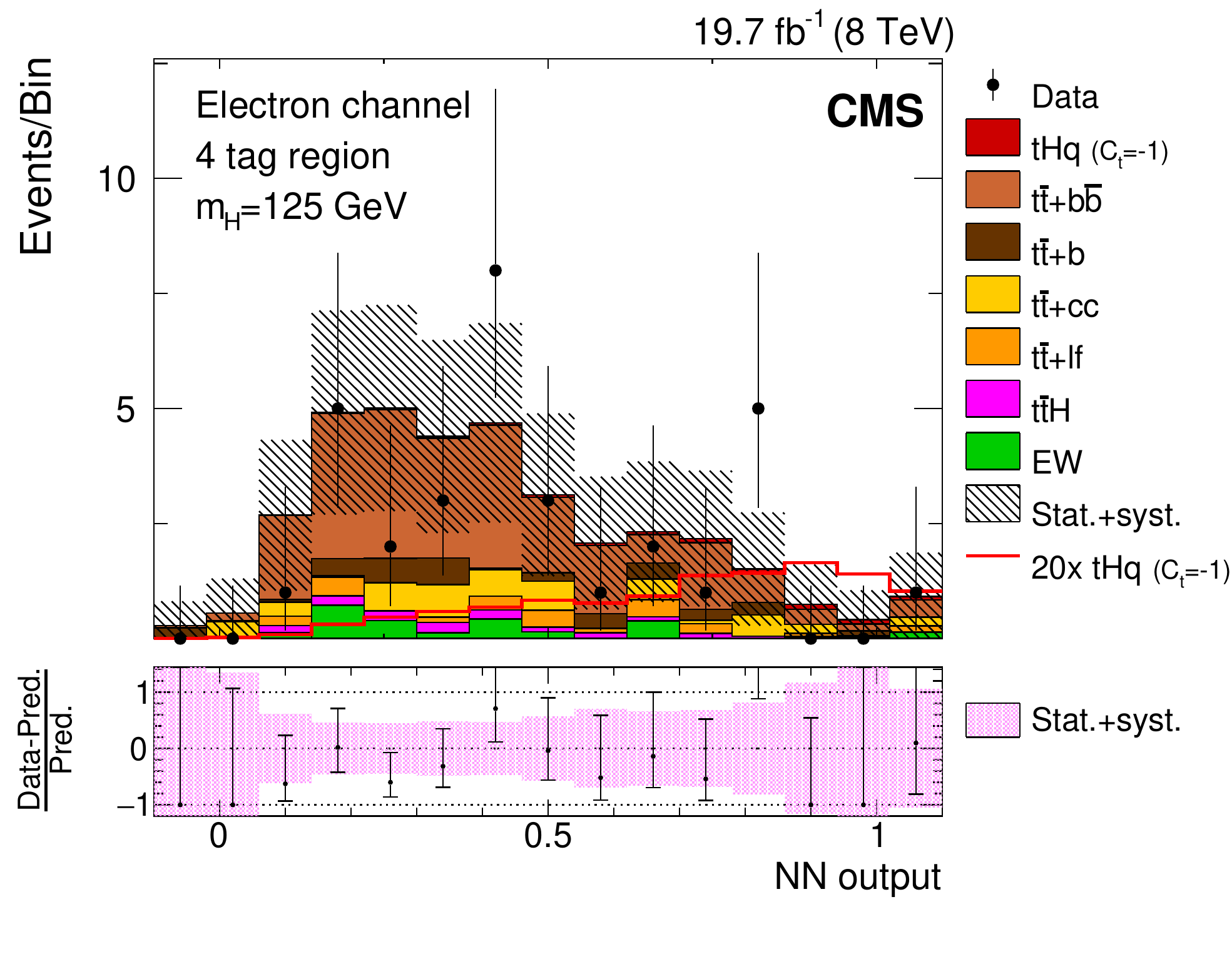}
    \includegraphics[scale=0.30]{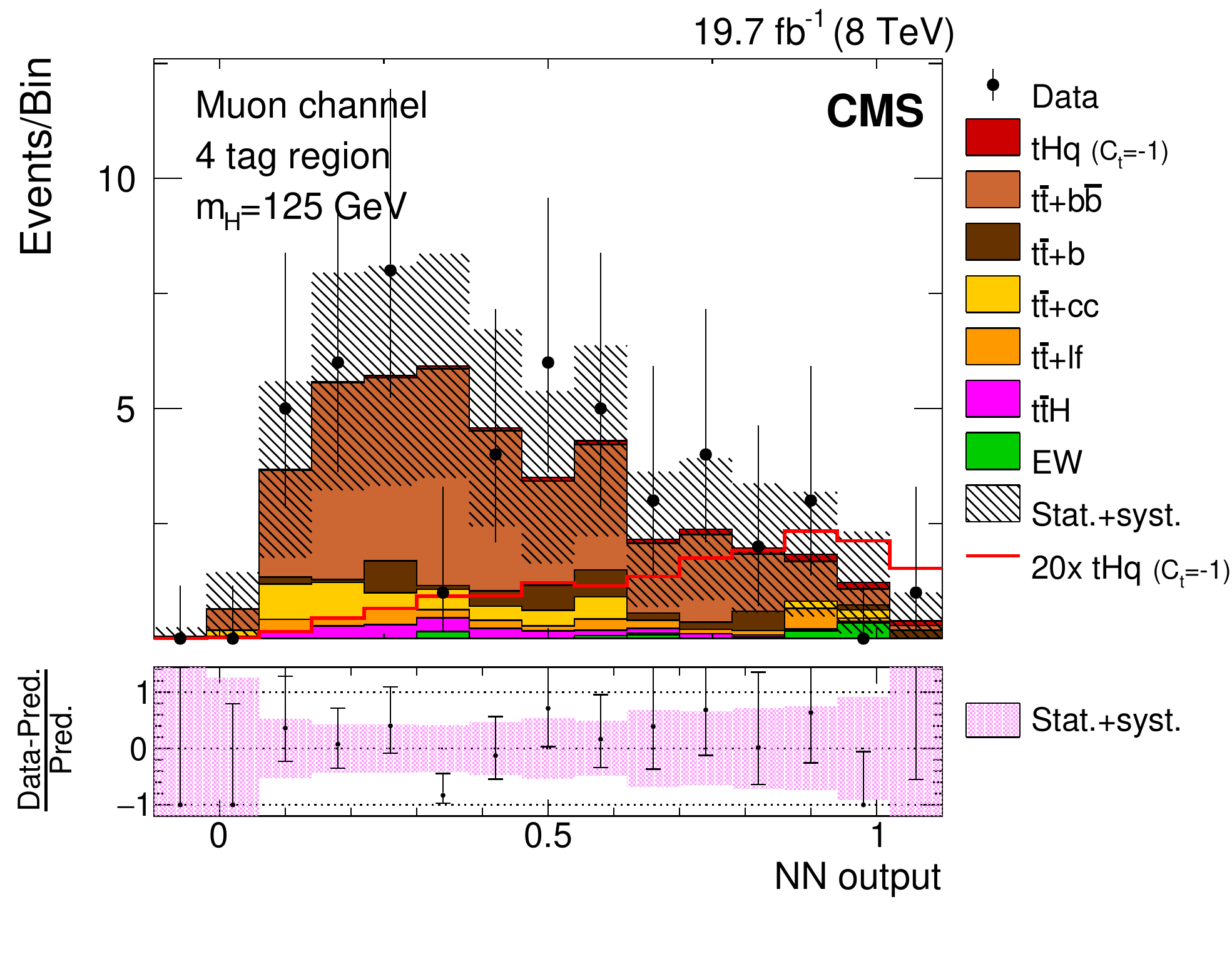}\\
    \caption{Distributions of the multivariate discriminator output for the $H \to b\bar{b}$
      channel for the three-tag electron and muon samples (upper row) and
      the four-tag electron and muon samples (lower row) for the selected
      data samples (points) and expected contributions from each physics
      process (histograms).  The histogram normalizations are set by the
      result of a maximum likelihood fit.  ``EW'' indicates electroweak
      backgrounds: single top, $W$/$Z$ plus jets, and di- and tri-boson
      production.  The line shows the expected contribution from the $tHq$
      process with $C_t = -1$ multiplied by the factor indicated in the
      legend.  In the box below each distribution, the ratio of the
      observed and predicted event yields is shown. The shaded band
      represents the post-fit systematic and statistical uncertainties.}
\label{fig:bbbar}
\end{figure}

Figure~\ref{fig:bbbar} shows the event-discriminator distributions for four
subsamples that are considered (3$b$ and 4$b$, with $e$ and $\mu$ as the
leptons), and the results of a fit that are used to set an upper limit on
the $tHq$ production rate.  The expected upper limit is
$5.2^{+2.1}_{-1.7} \times \sigma_{tHq}(C_t = -1)$ at the 95\% confidence level,
and the observed limit is $7.6 \times\sigma_{tHq}(C_t = -1)$.  The largest
systematic uncertainties come from the $t\bar{t}$ modeling and the $b$-tag
efficiencies and mistag rates.

\section{Combined results}

\begin{figure}[htb]
\centering
    \includegraphics[scale=0.30]{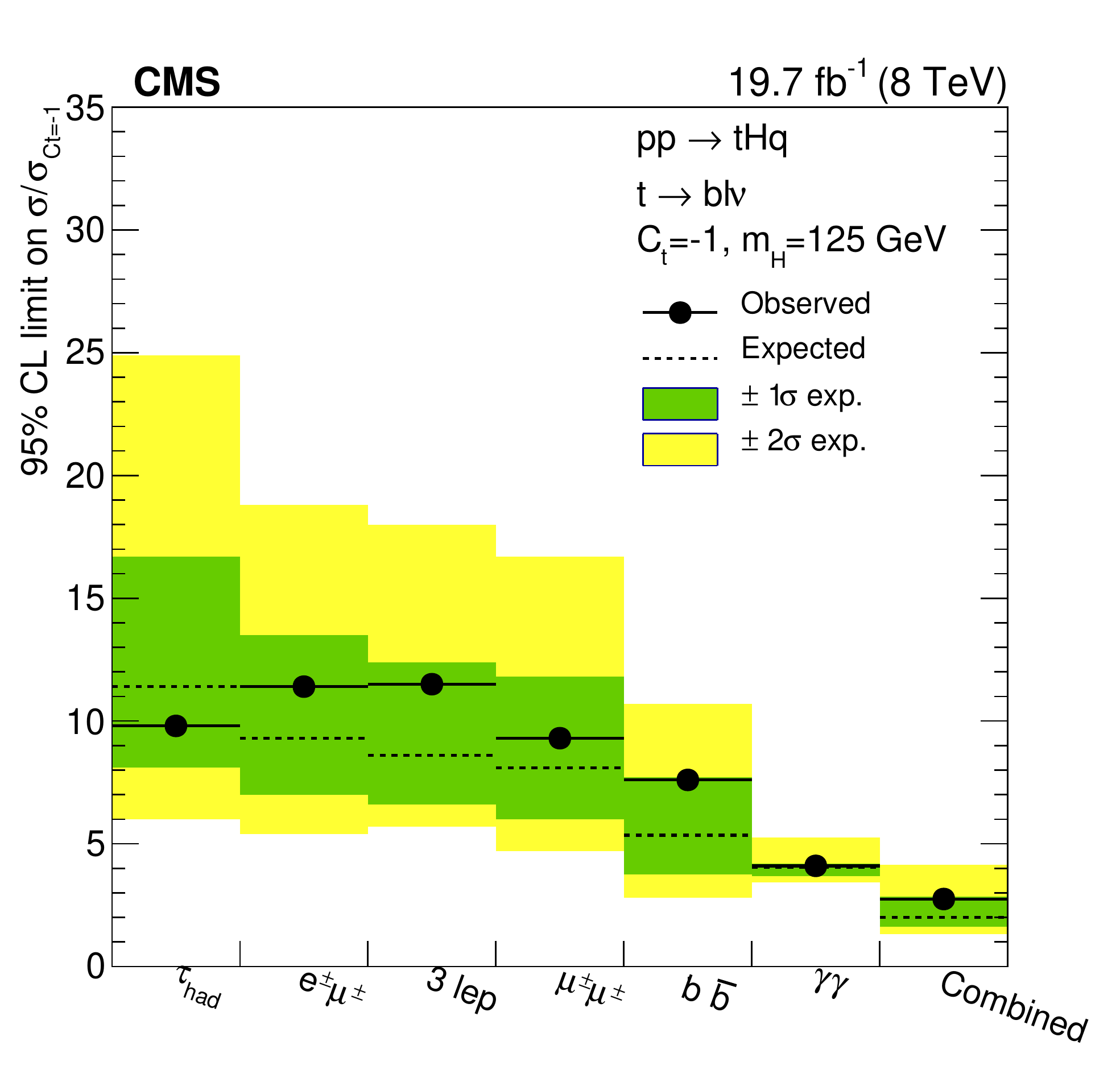}
    \includegraphics[scale=0.30]{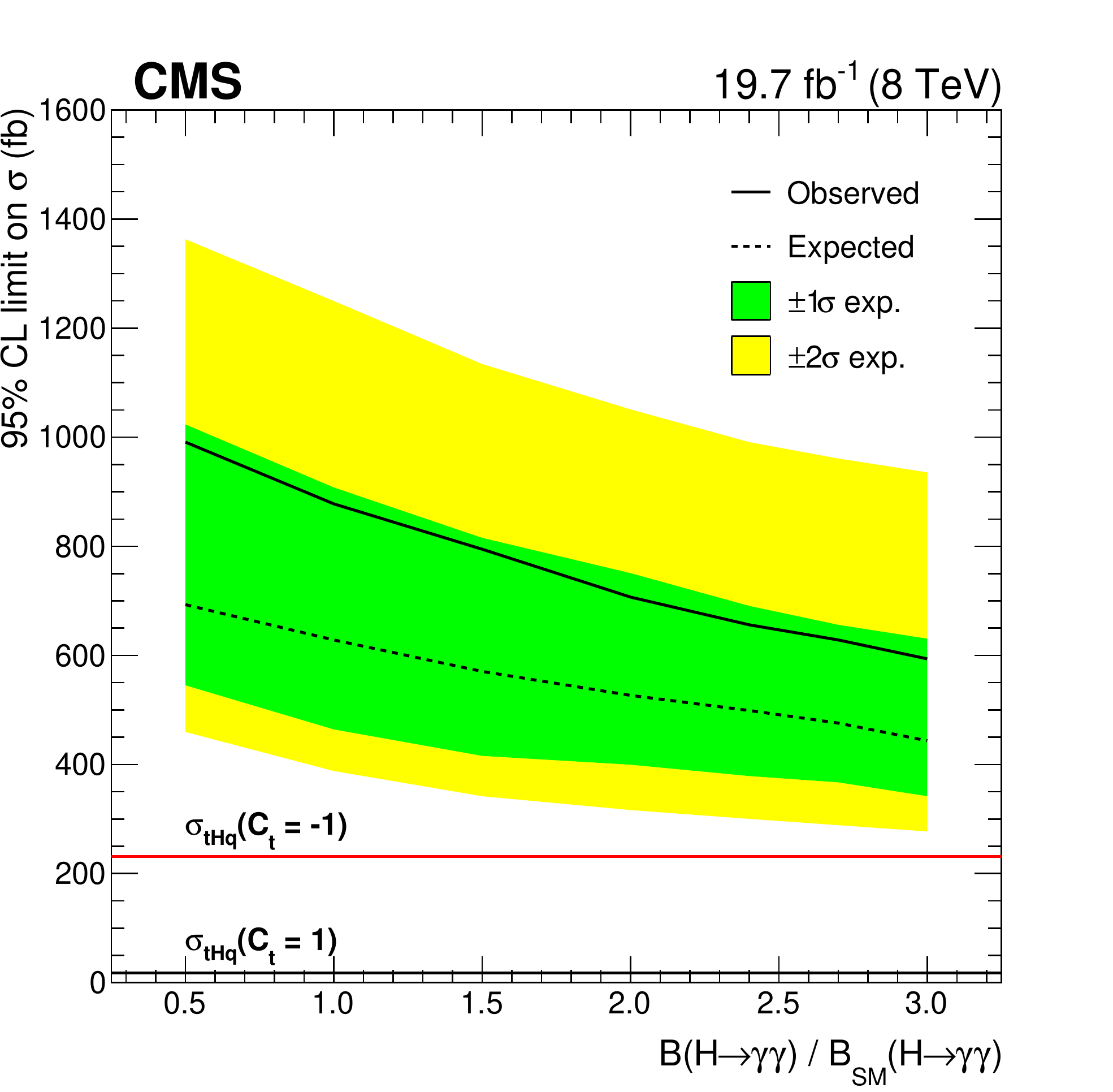}
    \caption{Left: The 95\% CL upper limits on the excess event yields
      predicted by the enhanced $tHq$ cross section and Higgs boson to
      diphoton branching fraction for $C_t=-1$. The limits are normalized
      to the $C_t=-1$ predictions and are shown for each analysis channel,
      and combined. The black solid and dotted lines show the observed and
      background-only expected limits, respectively. The $1\sigma$ and
      $2\sigma$ bands represent the 1 and 2~standard deviation
      uncertainties on the expected limits.  Right: The 95\% CL upper
      limits on the $tHq$ production cross section as a function of the
      assumed Higgs boson to diphoton branching fraction. The black solid
      and dotted lines show the observed and background-only expected
      limits, respectively. The $1\sigma$ and $2\sigma$ bands represent the
      1 and 2~standard deviation uncertainties on the expected limits.  The
      red horizontal line shows the predicted $tHq$ cross section for the
      SM Higgs boson with $m_{H}$ = 125~GeV in the $C_t=-1$ scenario,
      while the black horizontal line shows the predicted $tHq$ cross
      section in the SM (i.e., $C_t=+1$).}
\label{fig:limits}
\end{figure}

Table~\ref{tab:results} summarizes all of the limits on $tHq$ production
described so far.  All of the information can be put into a single
likelihood fit to obtain a single limit for all four searches.  The fit has
many bins and many free parameters to describe the systematic
uncertainties; careful attention is paid to correlations amongst the
uncertainties.  Two different approaches are taken.  One is to test
sensitivity to $C_t = -1$ and count any enhancement to $tHq$ production and
the $H \to \gamma\gamma$ decay rate as signal.  Under that assumption,
the expected limit is $2.0^{+0.8}_{-1.2} \times \sigma_{tHq}(C_t = -1)$ at 95\% confidence
level, and the observed limit is $2.8 \times\sigma_{tHq}(C_t = -1)$.  These
results are displayed in the left panel of Figure~\ref{fig:limits}.  An
alternative approach is to quote a limit on $\sigma_{tHq}$ as a function of
$B(H \to \gamma\gamma)$.  That result is also shown in Figure~\ref{fig:limits}.

\begin{table}[t]
\begin{center}
\begin{tabular}{c|cccc}  \hline
$\sigma_{95\%}/\sigma_{tHq}(C_t = -1)$ & $H \to \gamma \gamma$ & $H \to
                                                                WW$/$\tau\tau$
  & $H \to \tau_{\mathrm{had}}\tau_\ell$ & $H \to b\bar{b}$ \\\hline
Expected & 4.1 & 5.0 & 11.4 & 5.4 \\
Observed & 4.1 & 6.7 & 9.8 & 7.6 \\\hline
\end{tabular}
\caption{Summary of expected and observed values of $\sigma_{95\%}/\sigma_{tHq}(C_t = -1)$.}
\label{tab:results}
\end{center}
\end{table}

\section{Conclusions and outlook}
The $tHq$ production rate is sensitive to the sign of the top Yukawa
coupling and other new physics.  The CMS Collaboration has now completed searches for $tHq$
in four different final states; the $H \to \tau_{\mathrm{had}}\tau_\ell$
channel was presented for the first time at this conference.  A combination
of the four searches, also new for this conference, yields a cross section
limit of $\sigma < 2.8 \times \sigma_{tHq}(C_t = -1)$ at 95\% confidence
level with the 19.7~fb$^{-1}$ dataset of the 8~TeV LHC run.  A paper
describing all of these results was recently submitted for
publication~\cite{bib:paper}.  The measurement is not yet sensitive to the
anomalous production of $tHq$, but the prospects for Run 2 are quite good.
$\sigma_{tHq}(C_t = -1)$ is four times larger at 13~TeV, and there should
be enough LHC data in 2016 to exclude (or perhaps observe?) the $C_t = -1$
hypothesis.  Beyond that, it will be possible to set limits in the Higgs
fermion coupling-vector coupling plane, have sensitivity to Higgs-mediated
FCNC processes with $q = u,c$ and more.  There some very interesting
opportunities ahead in these searches.

\Acknowledgments I thank my many colleagues on CMS: the 3000 scientists and
engineers who built and operate the experiment, and the subset of them who
worked specifically on these results.  I also thank the leaders of the CMS
physics organization who allowed me to make the first presentation of some
of these results.  My work on CMS is supported by the National Science
Foundation through award NSF-1306040.

\end{document}